# Influence of Hydrogen on Dislocation Relaxation in BCC Iron: Atomistic Mechanisms and Implications

Sanjay Manda[1,$], Madhur Gupta[1,$], Saurabh Kumar[1], Junaid Akhter[1], P. J. Guruprasad[2], Indradev Samajdar[1,*], and Ajay S. Panwar[1,*]

[1] Department of Metallurgical Engineering and Materials Science, Indian Institute of Technology Bombay, Mumbai, 400076, India.

[2] Department of Aerospace Engineering, Indian Institute of Technology Bombay, Mumbai, 400076, India.

[$] Joint first co-authors.

## Abstract

In this study, the influence of pure dislocation and hydrogen-dislocation interactions on anelastic response or internal friction relaxation peaks in bcc-iron was investigated. These relaxations are primarily governed by thermally activated kink nucleation and kink migration events. An atomistic multiscale framework, coupling molecular dynamics (MD) and kinetic Monte Carlo (KMC) simulations, was developed to investigate the underlying atomistic mechanisms behind dislocation-relaxation peaks. MD simulations revealed that the presence of hydrogen atoms near the dislocation core facilitates the kink nucleation process by reducing the nucleation barrier while enhancing the barrier for dislocation migration. The KMC model captured Snoek-Köster peaks arising from the Cottrell atmosphere formed by hydrogen atoms and clusters around the dislocation core, providing insights into the atomistic mechanisms controlling these relaxations. Furthermore, the proposed computational scheme elucidated a unique linear relationship between hydrogen content and the internal friction loss factor, $\tan \delta$, offering a methodology for hydrogen detection and quantification.

**Keywords:** Hydrogen, Dislocation, Internal Friction, Molecular Dynamics, Kinetic Monte Carlo.

*Corresponding authors: panwar@iitb.ac.in, indra@iitb.ac.in.

## I. INTRODUCTION

Internal friction measurements have been utilized to characterize the anelastic (or viscoelastic) nature of metallic materials since the middle of the 20[th] century [1–7]. Internal friction can play a significant role in the geology field by offering valuable insights into the Earth's defect structure, the mechanisms underlying seismic wave attenuation, and their correlation with earthquakes [8,9]. The anelastic or time-dependent behavior of the material can lead to a dissipation of mechanical energy [1]. This dissipation could arise from various microstructural sites such as solute atoms [10–13], dislocations [14–16], and grain boundaries [17,18]. These contributing sources have characteristic frequency and temperature [7]. The present study is designed to investigate various aspects of dislocation relaxation or dislocation damping. In this context, the role of deformation, known to introduce dislocations, on internal friction response has been explored extensively [3,6,15,19]. Snoek [20], also reported the presence of a high temperature peak beside a room temperature peak (known as the Snoek peak) during the deformation. This was attributed to the long-range diffusion of solute atoms in deformation-induced macroscopic stress fields, which is similar to the well-known Gorsky effect [20]. Later, Kê [21] confirmed the presence of such a peak and called it a cold-work peak. Subsequently, elaborative studies from Köster [22–24] revealed that the activation energy associated with the cold-work peak was higher than that of the Snoek peak [23]. Later, Nowick and Berry [2] summarized all the studies related to the cold-work peak and termed it the 'Snoek-Köster peak'. Since the inception of this peak, multiple theories have been proposed to explain the mechanism behind this relaxation [4,5,15,16]. Dislocation relaxation in the presence of Cottrell atmosphere formed by solute atoms was regarded as a major mechanism behind the SK peak [4,25,26].

Since the coining of the SK peak, two main theories have emerged that explain the dislocation relaxation phenomena and the influence of the solute atoms [27–31]. For instance, Schoeck [27,28] formulated a string model to explain the anelastic strain response of the dislocation, believing that the bowing of the dislocation, which is constrained by the line tension, is the key mechanism behind the SK peak. In contrast, Seeger [29–31] argued that dislocation moves, between two Peierls valleys, through kink-pair formation. Since dislocation segments have a large formation energy associated with them, the kink-pair theory proposed by Seeger, which supports an energetically favorable approach, was adopted in subsequent studies [1,2,5,25]. In general, three types of internal friction peaks are known to arise from such dislocation relaxation processes [1,32,33]. Firstly, intrinsic dislocation relaxation can give rise to two distinct peaks, known as the α and γ peaks, which originate from pure edge and screw dislocations, respectively. These are collectively referred to as Bordoni peaks [1]. On the other hand, dislocation relaxation in the presence of solute atoms (C, H, and N) leads to the emergence of the SK peak [1,32,33]. In this work, we emphasized on exploring the science behind the $\gamma$ peak and hydrogen promoted SK peak, popularly known as the SK(H) peak. It should be noted that any such relaxation occurs through the diffusive motion of the dislocations, which is governed by the activation energy required for such transitions via kink nucleation and migration [34–37]. The variations in activation energy with imposed stress through MD simulations are very well documented [34–36,38]. However, the impact of hydrogen on the activation energy barrier for dislocation glide has not been explored much. Therefore, the current study is focused on evaluating the role of hydrogen and applied stress on kink-nucleation and migration processes. Moreover, the details about the underlying mechanism responsible for different dislocation-related peaks, ($\alpha, \gamma,$ and SK(H)) remain unexplored. Juan et al. [32,39] have attempted to explain these peaks through internal friction experiments on a pure iron sample subjected to plastic deformation and hydrogenation, as shown in Supplementary Figure S1. A theoretical model proposed by Juan et al. [32] qualitatively presents the internal friction relaxation in iron arising from dislocation-hydrogen interactions. The proposed model [30] holds solute drag responsible for SK(H) relaxation. However, these measurements [32,39] provided only qualitative information, and the fundamental principles were not discussed. This gap serves as the motivation for the present work, which aims to bring out the atomistic insight related to such relaxation phenomena. In another study by Jiang et al. [40], a small modification to Seeger's model [30,31] was proposed by incorporating the long-range ordering of solutes, especially hydrogen, owing to its higher diffusivity. The simultaneous movement of hydrogen and dislocations was indirectly proposed as a key mechanism governing the SK peak. To provide more insights, the different aspects of hydrogen-dislocation interactions have also been explored in this work.

It has been reported that most real materials are not fully elastic [1,8,32]; they have both elastic and anelastic components. Such a standard linear solid behavior can be effectively described physically using a spring-Voigt model, as illustrated in the Supplementary Information Fig. S2. The nature of anelastic response is governed by a combination of dashpot plus spring behavior. This behavior can be mathematically formulated as,

$$\sigma = \sigma_0 \, e^{i\omega t} \tag{1}$$

$$\varepsilon = \varepsilon_0 e^{i(\omega t - \delta)} = (\varepsilon' - i\varepsilon'') \, e^{i\omega t} \tag{2}$$

$$E^* = \sigma/\varepsilon = (E' - iE'') \tag{3}$$

$$Q^{-1} = \frac{E''}{E'} = \frac{\varepsilon''}{\varepsilon'} = \frac{\Delta W}{2\pi W} = \tan \delta \tag{4}$$

The term $\delta$ denotes the phase lag between input stress ($\sigma$) and output strain ($\varepsilon$), developing due to anelastic relaxation. Subsequently, both the strain ($\varepsilon$) and complex elastic modulus ($E^*$) can be separated into their elastic ($\varepsilon'$, $E'$) and anelastic ($\varepsilon''$, $E''$) components. The internal friction loss factor ($Q^{-1}$ or $\tan\delta$) can be estimated as the ratio of energy dissipated ($\Delta W$) to the stored energy ($W$). Alternatively, it can be expressed as the ratio of the loss modulus ($E''$) to the storage modulus ($E'$). In addition, the characteristic time scale ($\tau$) of the activated process(es), which play a significant role in energy dissipation, can be given as,

$$\tau = \tau_0 \exp\left(\frac{G}{kT}\right) \quad (5)$$

where $G$ is the activation energy associated with the process. Subsequently, the internal friction peak can also be presented as,

$$Q^{-1} = 2\, Q_{max.}^{-1} \frac{\omega\tau}{(1 + \omega^2 \tau^2)} \quad (6)$$

Eq. 6 results in a bell-shaped peak or a Debye peak, whose characteristics are governed by the applied frequency ($\omega$) and the characteristic time scale ($\tau$). When these two parameters are in resonance, i.e., $\omega\tau = 1$, energy dissipation is highest, corresponding to the peak height ($Q_{max}^{-1}$). Therefore, both the externally imposed time scales and the internal relaxation time scales are crucial, and their importance is deliberated later in this work.

In recent times, hydrogen has received significant attention as an alternative fuel due to its abundance and environmentally friendly characteristics [41,42]. However, two major problems are associated with hydrogen. One serious concern is that hydrogen degrades the mechanical performance of the materials, which is termed 'hydrogen embrittlement' [41,42]. This problem has been thoroughly investigated in our previous study [43] and by other researchers [44–46]. On the other hand, the detection and quantification of hydrogen has been recognized as another concern by the research community [47]. The small size and faster diffusion of hydrogen make their detection more challenging. In this context, internal friction measurements, which are sensitive to several microstructural variables, can be employed as a potential technique for hydrogen detection and quantification. In addition, it is possible to characterize the internal friction peaks with respect to the type of hydrogen, atom versus cluster. Many scientific questions regarding hydrogen remain unanswered, which could potentially be addressed through targeted investigations linking specific microstructural variables with internal friction measurements. This study aims to address some of these questions through multiscale atomistic simulations combining MD-KMC methods.

## II. SIMULATION METHODOLOGY

The present study aims to simulate the dislocation movement and associated internal friction relaxation processes. This was achieved by using a combination of MD and KMC methods. The details about the implementation of these methods are provided below.

### A. Molecular Dynamics Simulations

The MD simulations were adopted to compute the activation energies for dislocation glide [48]. The initial and final configurations of screw dislocation with burgers vector $b = \frac{1}{2}[111]$ was generated using open access Atomsk$^{TM}$ package [49]. The principal axes x, y and z are oriented along the $[\bar{1}2\bar{1}]$, $[\bar{1}01]$ and $[111]$, respectively, having the box dimensions as follows: $L_x = 138$ Å; $L_y = 107$ Å; $L_z = 132$ Å, see Supplementary Information Fig. S3. The stable position of the dislocation is referred to as the Peierls valley. On application of external stress, the screw dislocation glides to the nearest Peierls valley along the slip direction <112> and the closed-paced plane {110}.

The transition between Peierls valleys depends on the height of the valley, which is also known as the activation energy barrier. These barriers were computed through the nudged elastic band (NEB) simulations, which were performed using the Large-scale Atomic/Molecular Massively Parallel Simulator (LAMMPS) package [50] and the embedded atom method (EAM) method developed by Proville et al. [51]. Further details regarding the implementation of the NEB method are available in the Supplementary Information section S1 and elsewhere [52,53]. Present simulations were carried out over a range of imposed shear stress values; the atomic configuration is depicted in Supplementary Information Fig. S2. Additionally, MD simulations were expanded to incorporate the influence of hydrogen concentration and its distributions on dislocation glide behavior. A hybrid potential was employed, combining the EAM potential by Proville et al. [41] for Fe-Fe interactions and the EAM potential by Wen et al. [44] for Fe-H and H-H interactions. Details on the hybrid potential and its benefits are provided in the Supplementary Information section S2.

### B. Kinetic Monte Carlo Simulations

The activation energies from MD simulations were used as input for our in-house KMC model. These KMC simulations provided relevant information regarding the long-range diffusion of dislocation [54] and their interactions with hydrogen. The dislocation glide is controlled by kink nucleation and migration events. Further, these two events can occur in both forward and backward directions. These four primary events are shown in Supplementary Information Fig. S4a. The screw dislocation is discretized into smaller segments of length, $l_{seg} = b$ along the [111] direction. These segments can move forward or backward, representing kink nucleation and migration between consecutive Peierls valleys separated by a distance $\lambda$ along the [$\bar{1}2\bar{1}$] direction. For a stable kink, the minimum width is approximately $w_{kn} = 25\ b$, and each migration step occurs over a distance of $b$. The transition probability of these events will be determined by the transition rate $(r)$, that can be expressed as,

$$r = v_o\ exp\left(-\frac{\Delta H}{kT}\right) \quad (7)$$

where $v_o$ is the attempt frequency. For the present study, $v_o$ for kink nucleation was taken $2 \times 10^{10}\ s^{-1}$ while $v_o = 5 \times 10^{11}\ s^{-1}$ was considered kink migration as per literature data from Shinzato et al. [34]. The attempt frequency for kink migration is approximately 25 times that of kink nucleation, and this relationship is consistent with the characteristic length scales of these events. Kink nucleation involves the formation of a new segment, having a length of ~ $25b$, whereas these kinks can migrate by a much shorter distance of only ~1b. $\Delta H$ is defined as an activation enthalpy or diffusion barrier for the transition. $\Delta H$ values were computed with NEB simulations. Among multiple possible events, a particular event was chosen based on the n-fold way algorithm [55]. These protocols are elaborated in Supplementary Information Fig. S4b. After each successful transition, simulation time $(t)$, which represents real time, is updated as per the following,

$$t = t_i - \frac{\log \rho}{R} \quad (8)$$

where $t_i$ represents the time at the start of a particular event, $R$ is the cumulative rate, and $\rho \in [0\ 1)$ is a random number, which signify the probabilistic nature of the transitions. These KMC simulations will provide the information related to dislocation evolution, as a function of stress $(\sigma)$ at different temperatures $(T)$ and time intervals $(t)$. The estimation of the resultant strain $(\varepsilon)$ is slightly tricky, and was calculated using the following relation,

$$\varepsilon = \frac{\sum_{i=1}^{N}(x_i)_t - (x_i)_0}{N} \quad (9)$$

where $(x_i)_t$ represents $x$ coordinate of $i^{th}$ atom at a simulation time, t while $(x_i)_0$ is $x$ coordinate of $i^{th}$ atom at the start of the simulation. $N$ denotes number of Fe atoms forming a screw dislocation. Since dislocation moves along its glide direction ($x$ axis || [$\bar{1}2\bar{1}$]), displacement of atoms along the $x$ axis has been considered for strain calculations. It should

be noted that Eq. 9 provides the average strain of the dislocation line. The overall cyclic stress ($\sigma(t)$) and strain ($\varepsilon(t)$) values can be expressed as follows,

$$\sigma(t) = \sigma_0 \sin(\omega t) \tag{10}$$

$$\varepsilon(t) = \varepsilon_0 \sin(\omega t + \delta) \tag{11}$$

The phase lag, $\delta$ has developed due to the anelastic nature of dislocation evolution. This can lead to several internal friction peaks. These peaks and associated atomistic mechanisms have been deliberated in the subsequent sections. It should be noted that the present simulations were performed at an imposed frequency of $10^4$ Hz over a temperature range of $100 - 400$ K.

## III. RESULTS AND DISCUSSION

### A. Activation Energy Barrier for Dislocation Glide

The simulation setup containing a single screw dislocation is shown in Fig. 1a. The NEB method was employed to estimate the activation energy or diffusion barrier for dislocation glide between appropriate sites. The resultant enthalpy profile, for the dislocation glide in the absence of any external stress, is shown in Fig. 1b. Representative dislocation configurations illustrating kink nucleation and migration events are also presented. Notably, a critical length of approximately $25b$ is required for stable kink nucleation, as reported in the literature [36,56]. Subsequently, these kinks can recombine or annihilate depending on their configurations and the prevailing testing conditions. Fig. 1c shows enthalpy profiles at different applied stress values. It is clear that external stress reduces the height of the energy or enthalpy landscape, which indicates stress-assisted motion of the dislocation. The outcomes of the NEB simulations provide the internal energy of the system, which includes the contributions from potential and kinetic energy components. The contribution of mechanical work in terms of applied stress was incorporated to calculate the enthalpy ($H_i$) of the system as follows,

$$H_i = U_i - \sigma b L \theta \tag{12}$$

where $U_i$ is the internal energy, $\sigma$ is the applied stress, $L$ is the length of dislocation, and $\theta$ is the reaction coordinate in distance units. The term ($\sigma b L \theta$) represents the mechanical energy of the system. Moreover, our diffusion barrier values show a very good quantitative match with the simulation results from Shinzato et al. [34] and Narayan et al. [38], as shown in Supplementary Fig. S5.

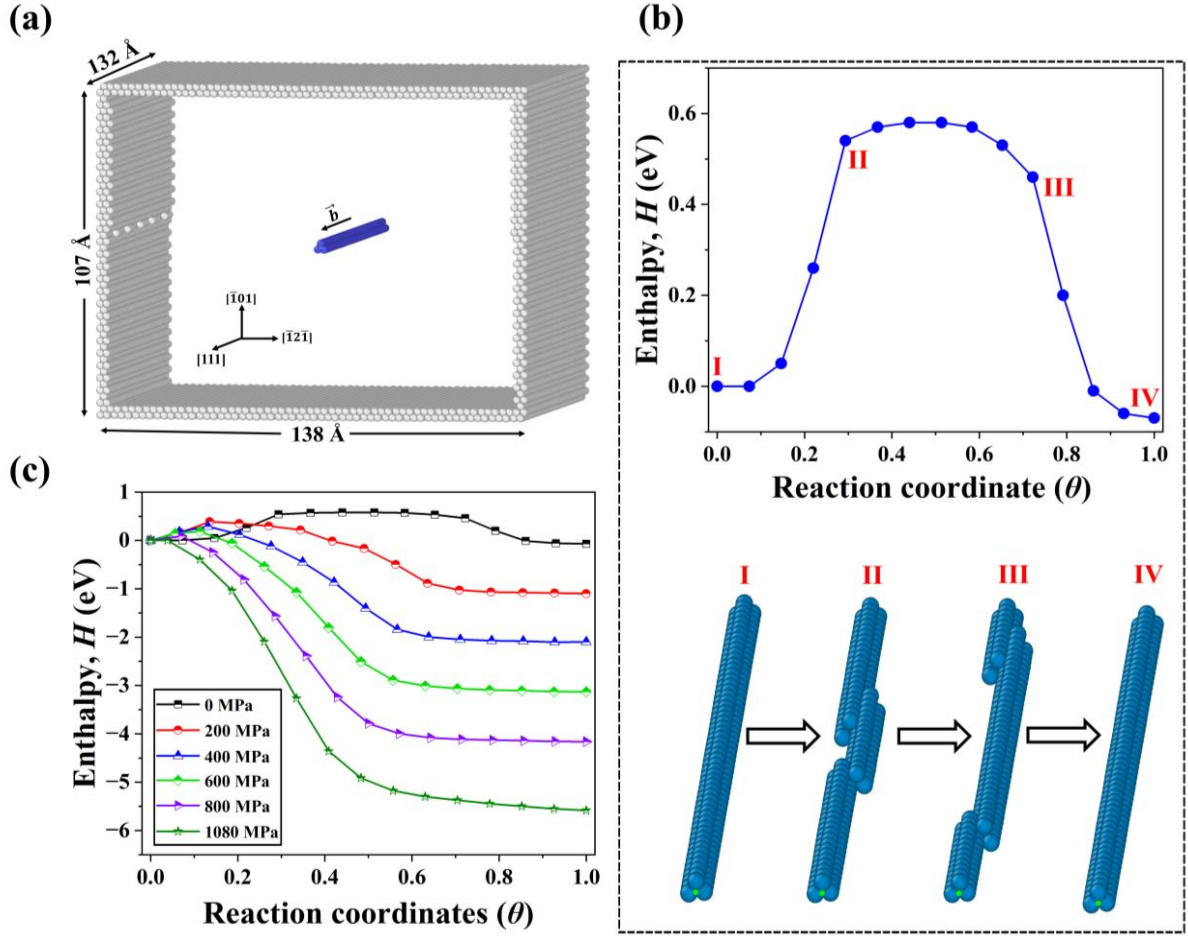

FIG. 1. (a) MD simulation setup containing one screw dislocation in the center. (b) The minimum energy path for screw dislocation estimated through NEB and the corresponding replicas representing the atomic configurations of dislocation along the minimum energy path. (c) Estimating energy landscape (enthalpy versus reaction coordinate) for different imposed shear stress values.

**B. Hydrogen-Dislocation Interactions**

The movement of dislocations is known to be affected by the presence of solute atoms in addition to temperature and applied stress [34]. This work examines the influence of the Cottrell atmosphere formed by hydrogen atoms on dislocation migration behavior. In this regard, the MD simulations were conducted, with various EAM potentials, to estimate the nucleation and migration barriers for the Fe-H system. However, none of these potentials predicted proper diffusion barrier values due to the absence of appropriate EAM potentials. This issue has been elaborated in the Supplementary Information section S2. To address this problem, an alternate method, reported by Shinzato et al. [56] for the Fe-Si system, was adopted. In this method, the potential energy contributions from Fe-H interactions were computed through simple molecular statics simulations and termed interaction energy. This method presents a unique methodology, that can be utilized to estimate diffusion barriers for more complex alloy systems, where suitable interatomic potentials are lacking. This approach was applied to the Fe-H system for the first time, and it worked very well, as described below.

In this scheme, hydrogen atoms were schematically placed in energetically-stable tetrahedral sites at variable distances from the dislocation core as shown in Fig. 2a. Site-1 is located closest to the dislocation core, while site-5 is located at the farthest point from the dislocation. The figure also demonstrates the attractive nature of dislocation toward hydrogen atoms. Subsequently, the interaction energy ($\Delta E_{sol.}$) between the dislocation and hydrogen atoms was estimated, as per the following formulation,

$$\Delta E_{sol.} = E_1 + E_2 - E_3 - E_4 \tag{13}$$

where $E_1$ is the potential energy of the system with a screw dislocation and hydrogen atom, $E_2$ is the potential energy of an ideal system without any dislocation and hydrogen atom, $E_3$ belongs to a system with only a screw dislocation and $E_4$ represents the energy of an ideal system with only hydrogen atom. Among different EAM potentials, the hybrid potentials combining Wen's [57] and Proville's [51] EAM potentials yielded the best results, as evident from Fig. 2a. The hydrogen atom at site-1 made the major contribution to the interaction energy term due to the shorter distance. The effect of dislocation and hydrogen interaction is significant till site 3 and reaches a stagnant value. These values are very close to the literature values available from Wen et al. simulations [58]. Consequently, the total enthalpy incorporating the contribution of interaction energy was calculated as follows,

$$\Delta H_t = \Delta H_d + \Delta H_{sol.} \tag{14}$$

where $\Delta H_t$ is the overall enthalpy of the system, $\Delta H_d$ represents the enthalpy of the system containing only dislocations, and $\Delta H_{sol.}$ is the energy contribution from hydrogen-dislocation interactions. Subsequently, kink nucleation barriers were estimated in the presence of hydrogen atoms and clusters at various sites. The outcomes of these simulations are summarized in Fig. 2b and compared against pure Fe values. It was evident that hydrogen significantly reduces the barrier for kink nucleation. In particular, hydrogen atoms located near the core had the strongest influence on the nucleation barrier. The nucleation barrier was found to be the lowest for the hydrogen cluster, attributed to the combined effect. Similarly, the migration barrier was also calculated with the formulation, as described in the Supplementary Information section S3. The values for the migration barrier are shown in Fig. 2c. The migration barrier was found to be very negligible (~$10^{-5}$ eV) for the pure Fe system. On the other hand, the migration barrier increased significantly in the vicinity of the hydrogen atmosphere.

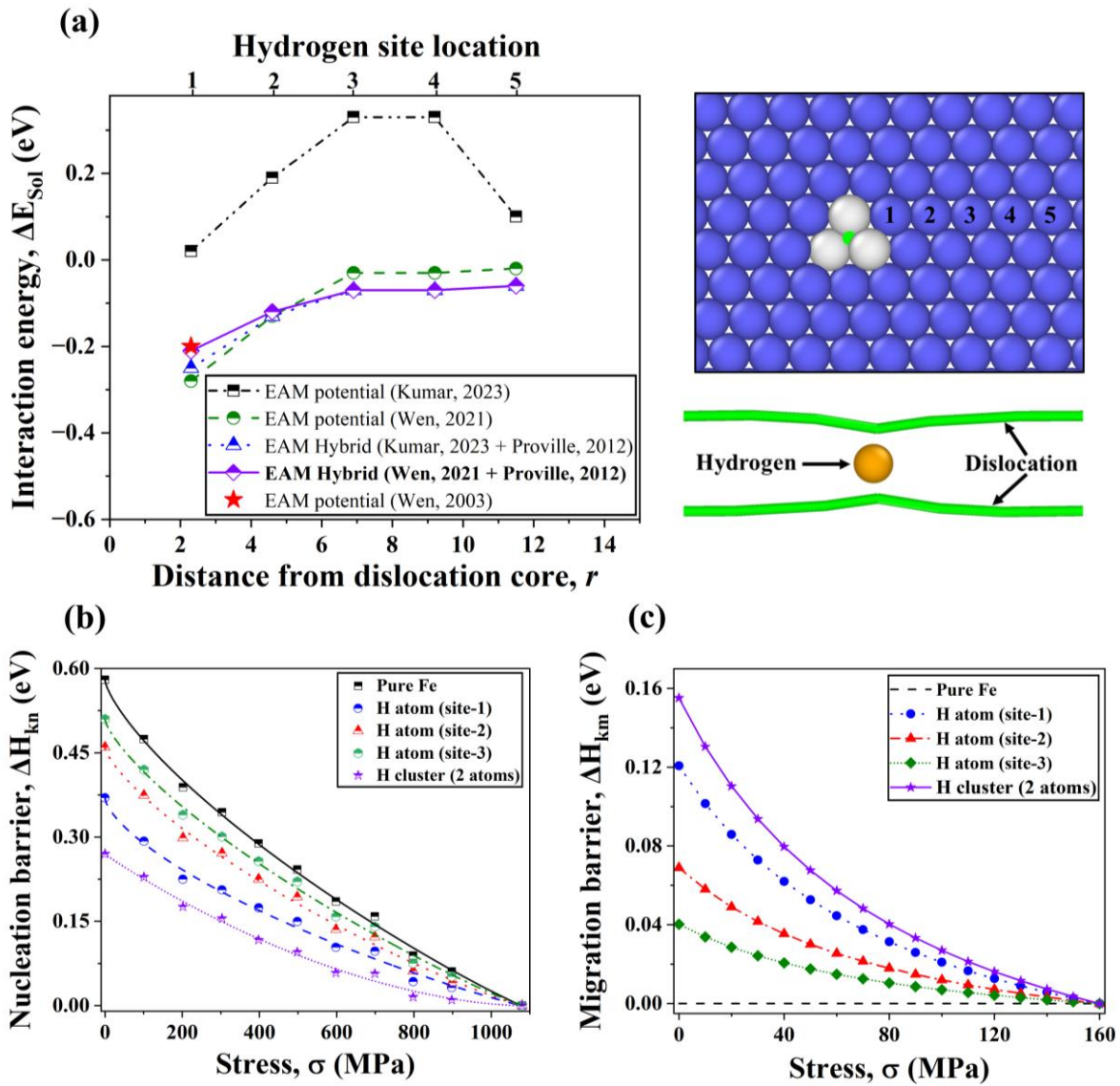

FIG. 2. (a) Estimation of hydrogen-dislocation interaction energy with different EAM potentials. Hydrogen atoms are placed at different locations from the core. (b) Nucleation barrier and (c) Migration barrier for dislocation movement in hydrogen atmosphere as a function of applied shear stress.

In summary, hydrogen promotes kink nucleation while hindering kink migration. The possible reasons for this behavior are deliberated later in the work. There have been a few studies that explored the dual role of hydrogen concentration with various solute atoms such as hydrogen, carbon, and nitrogen [59–62]. In such studies [59–62], the dislocation type, stress, temperature, and solute concentration were regarded as multiple factors deciding the overall nature of the dislocation movement. For hydrogen-related studies [59,60], the primary focus was on the overall influence of hydrogen content on dislocation glide. However, a systematic investigation into the individual effects of hydrogen on dislocation nucleation and migration barriers has not been conducted. This study aims to provide novel insights into the variations of nucleation and migration barriers in a hydrogen atmosphere. According to Fig. 2, the nucleation barrier is much higher than the migration barrier for the pure Fe system, the nucleation barrier decides the overall dislocation glide behavior. On the other hand, the nucleation barrier decreased while the migration barrier increased in the presence of a hydrogen atmosphere around the dislocation. Therefore, both nucleation and migration barriers will play a crucial role in dislocation movement, and they are sensitive to the hydrogen concentration. As a result, hydrogen concentration will decide the overall mobility of the dislocations. Overall, the present MD simulations provided estimates of the nucleation and migration barriers governing the dislocation glide process. Subsequently, these values were fed as input parameters for KMC simulations to simulate dislocation glide behavior and resulting internal friction peaks over a range of temperatures and timescales.

**C. KMC Simulations of Anelastic Dislocation Relaxation**

KMC simulations were conducted to model the atomistic interactions between hydrogen and dislocation, aiming to investigate their influence on the resulting internal friction response. Fig. 3a illustrates the framework of the present KMC model used to simulate the movement of a screw dislocation, governed by kink nucleation and kink migration events. Since the KMC method is based on the probabilistic nature of transitional events, a list of such events needs to be built beforehand, these events are: (i) forward kink nucleation, (ii) backward kink nucleation, (iii) forward kink migration, and (iv) backward kink migration. The nature of these events is decided by the direction of the applied stress and corresponding rates. The dimensional change during the nucleation event is very high ($w_{kn}$ = 25$b$) compared to the migration event ($l_{seg} = b$). Moreover, the transition rate of kink nucleation is much lower than the migration event. Therefore, kink nucleation events will play a major role in the overall anelastic relaxation behavior of dislocation relaxation. Subsequently, the impact of hydrogen on dislocation glide behavior has been evaluated. For this purpose, hydrogen atoms and clusters were schematically placed near the dislocation, as shown in Fig. 3b. Further, dislocation velocities were calculated over a range of applied stresses and temperatures. In this context, the time evolution of the screw dislocation under a constant stress of 100 MPa at 300 K is shown in Supplementary Information Fig. S7a, highlighting various stages of long-range diffusion. It was observed that the present simulations could maintain the connectivity between the various segments of the dislocation line and accurately capture the evolution of a straight dislocation into complex configurations. Moreover, Supplementary Information Fig. S7a also depicts the displacement-time evolution of the dislocation core. To ensure statistical reliability, we considered 20 independent simulations (Supplementary Information Fig. S7b), and by averaging the outcomes, we obtained the average dislocation velocity ($v_d$) as a slope between dislocation displacement and simulation time, which shows a good agreement with the existing literature by Shinzato et al. [34], as shown in Supplementary Information Fig. S7c. Supplementary Information Fig. S7d illustrates the variation of dislocation velocity under different applied stresses, demonstrating that the dislocation velocity increases with both temperature and applied stress.

More importantly, additional calculations were also carried out to understand the contributions from hydrogen atoms and clusters by introducing 0.05 at% hydrogen. To analyze their interactions with the dislocation, we considered two systems with the same overall hydrogen concentration: the first system had only hydrogen atoms, while the second system contained only hydrogen clusters around the dislocation. Notably, dislocation velocity was enhanced significantly in the vicinity of hydrogen atoms and clusters, as presented in Fig. 3c. Since most of the hydrogen (~90%) exists in the form of atoms, the change in dislocation velocity is quite large, corresponding to hydrogen atoms. These simulations of dislocation velocity demonstrate the potential of the present KMC setup in capturing dislocation glide behavior accurately and its implications towards real problems. In addition to dislocation glide, the role of hydrogen needs to be considered. It is to be noted that hydrogen, being smaller in size, diffuses quite readily and has a bigger impact on various properties of the real systems. Therefore, it is logical to incorporate both these effects to replicate a more realistic phenomenon. The corresponding rates of these two processes were calculated and are presented in Fig. 3d. It is clear that hydrogen, being a faster moving specie, diffuses at very high rates (~$10^{11} - 10^{15}\ s^{-1}$) while dislocation velocity is comparatively lower (~$10^4 - 10^7\ s^{-1}$). The difference in diffusion rates is quite significant, and capturing both hydrogen and dislocation diffusion within the same simulation time frame is impractical. Thus, it was assumed that the hydrogen atoms and clusters remain static, meaning they neither diffuse nor are dragged by the moving dislocation. In contrast, San Juan [32] proposed the dragging of hydrogen by moving dislocations under certain stress and temperature conditions. While this hypothesis holds a significant merit, it was not validated further. This phenomenon was not considered in the present study attributed to the computational limitations. Though hydrogen diffusion or drag effects were not considered, the role of hydrogen on dislocation movement was schematically evaluated. Subsequently, dislocation glide behavior in the presence of a cyclic stress was investigated, and the resulting cyclic strain was computed. The corresponding cyclic stress-strain response is presented in Fig. 4a for 10 such cycles.

Since kink nucleation and migration events occur at particular time scales, this led to time-dependent or anelastic behavior of the dislocation glide process. It was observed that the strain signal developed a phase lag of $\delta$. This delayed response or phase lag is compensated by energy dissipation within the system. The source of this phase lag lies in the rate sensitive nature of dislocation relaxation. It should be noted that Fig. 4a contained an average stress-strain response, which was obtained from multiple trajectories shown in Supplementary Information Fig. S8. The present simulations were then utilized to extract atomic-level information associated with the dislocation glide process. The mechanism associated with anelastic or time-dependent dislocation relaxation, which gives rise to the $\gamma$ peak, is illustrated in Fig. 4b. This is demonstrated as a series of selected snapshots of the dislocation glide event. These snapshots revealed the evolutions of screw dislocation line in the form of small segments governed by kink nucleation and migration events. It is also evident that dislocation movement is limited by the size and density of the kinks. Even when the stress cycle was reversed, the nature of the strain cycle, arising from dislocation movement, doesn't change immediately. It is known that dislocation movement is controlled by both kink nucleation and migration rates, and their time-scales are quite different than imposed time scales. Due to the time-sensitive nature of these events, a phase lag, $\delta$, develops between the input signal (stress) and the output signal (strain). The extent of this phase lag is responsible for the overall internal friction spectrum. This phase lag, $\delta$, results in an internal friction peak, which was simulated as $\tan \delta$ over different temperatures. Intrinsic dislocation relaxation peak appeared in the temperature range of 300-400 K, see Fig. 4c.

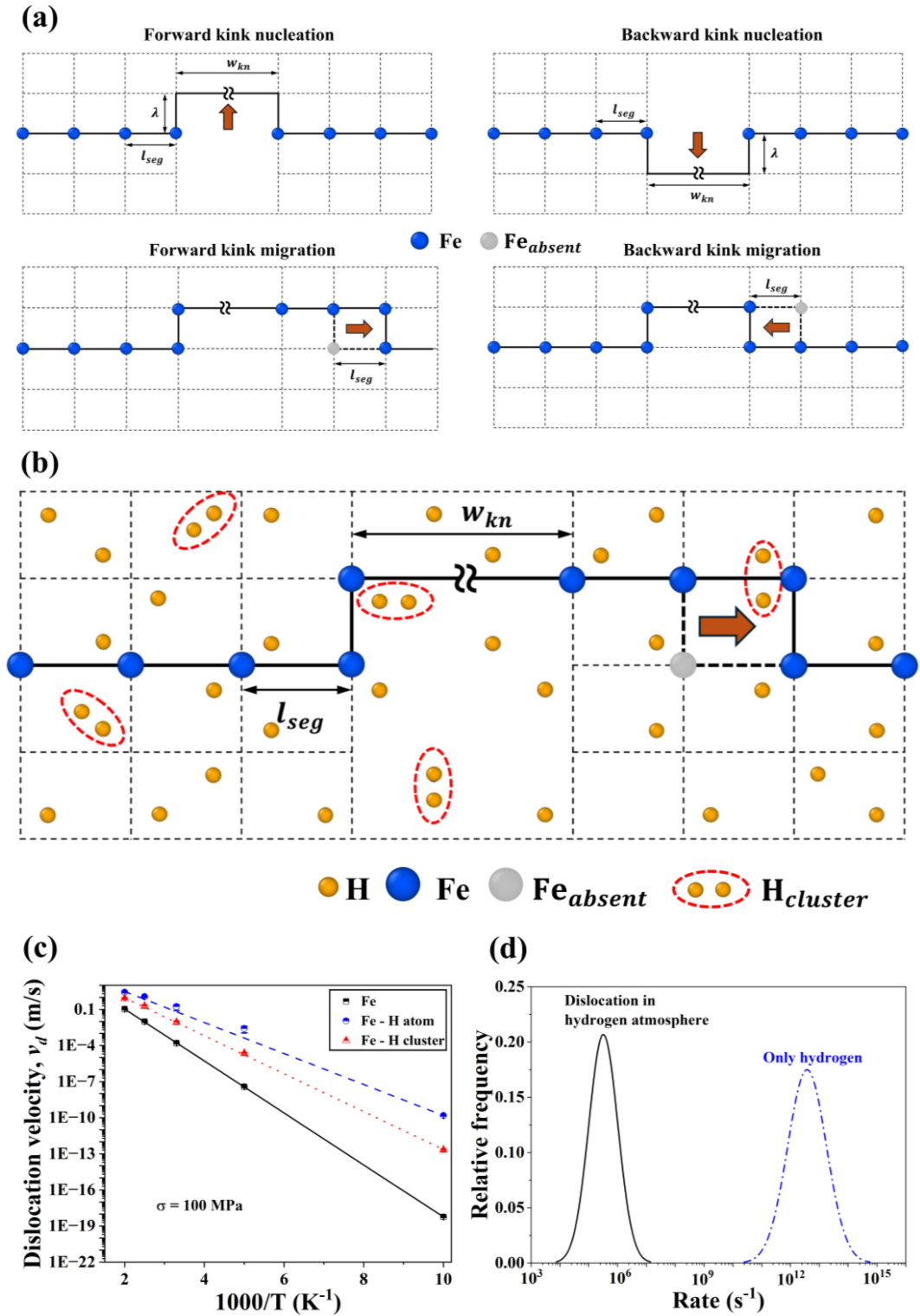

FIG. 3. (a) KMC setup demonstrating the movement of dislocation through kink nucleation migration events. (b) KMC setup showing the presence of a screw dislocation, hydrogen atoms and clusters. (c) Dislocation velocity plotted versus temperature for pure Fe, Fe-H (0.05 at% H) system. (d) Rate of hydrogen atom diffusion and dislocation glide process.

This peak is normally recognized as the $\gamma$ peak and appeared through thermally-activated kink nucleation and migration events at various locations of dislocation segments. The extent of dissipation will strongly depend on the similarities in the relaxation time of internal defects (dislocation in this case) and external or imposed time scales. The highest $\tan\delta$ is noticed when both scales are in resonance. This peak can be quantified in terms of peak position and height. The peak position is governed by the activation energy of the associated event, while the concentration of the defects will decide the magnitude of the peak.

Later, hydrogen was introduced in the simulation system in two forms: atoms and clusters. Unlike experiments, it was possible with our simulations to isolate the contributions of hydrogen atoms and clusters on dislocation relaxation phenomena. For this purpose, 0.05 at% hydrogen was introduced into the simulation box containing a screw dislocation. It was observed that most of the hydrogen (~90%) was present in the form of atoms, while the remaining hydrogen formed clusters. In the first set of simulations, the influence of hydrogen atoms on the dislocation relaxation peak was explored. It was noticed that the internal friction peak shifted to low temperatures due to hydrogen atom interactions with the dislocations. This peak is classified as SK(H)-I peak, as illustrated in Fig. 4c. The second set of simulations revealed another peak, SK(H)-II peak, arising from hydrogen clusters interactions with the screw dislocation, see Fig. 4c. Both forms of hydrogen, i.e., atom and clusters, led to an increase in $\tan\delta$ attributed to hindrance in the dislocation migration process in the hydrogen atmosphere. Notably, the peak height of SK(H)-I was more than SK(H)-II due to enhanced interactions between hydrogen atoms and dislocation segments. These simulations were performed by incorporating hydrogen atoms and clusters separately. However, hydrogen can co-exist in both forms in real scenarios. To bring this up, additional simulations were carried out in which hydrogen atoms and clusters were included simultaneously in the simulation system. The resultant internal friction spectrum is quite complex, as presented in Figure 4d. This spectrum was deconvoluted into two SK peaks, SK(H)-I and SK(H)-II. The deconvoluted peaks closely resemble the individual SK(H) peaks, shown in Fig. 4c.

Moreover, the time-dependent behavior of standard anelastic solid can be represented as follows [2],

$$A\dot{\bar{X}} + B\bar{X} = Cb\sigma + b\sigma \qquad (15)$$

here $B = M'_R b;\quad A/B = \tau_\sigma;\quad C = \tau_\varepsilon;\quad A/B = M'_U b$.

where $\tau_\sigma$ and $\tau_\varepsilon$ denotes the relaxation times at constant stress and strain, respectively, $M'_R$ and $M'_U$ represents the relaxed and unrelaxed modulus, respectively, $\sigma$ is the applied stress, and $\bar{X}$ is the average displacement of dislocation. The above relation can be extended to study the dislocation behavior in the presence of externally applied stress over the range of temperatures and time scales. Similarly, the theoretical model presented by Juan et al. [32,39] is quite relevant to the present work and is described below. The internal friction for the anelastic solid can be given by,

$$Q^{-1}(\omega) = \left(\frac{\xi}{6}\right)\Lambda(L_P^2 - L_H^2)\{\omega\tau/[1+(\omega\tau)^2]\} \qquad (16)$$

where $\xi$ is a coefficient of order 0.1, $L_P$ is the length of dislocation, $L_H$ is the length of dislocation between two hydrogen atoms, and $\tau$ is the relaxation time. In the absence of a hydrogen atmosphere, the whole dislocation segment length $L_P$ is responsible for anelastic relaxation and resultant internal friction losses. Whereas, in the presence of hydrogen, the dislocation is deduced to smaller dislocation segments of variable length, $L_H$, contributing as sources for anelastic relaxation. Since hydrogen creates more such sources of anelastic relaxation, the corresponding losses or peak height of SK(H) peaks are more than the height of $\gamma$ peak. Considering the assumption that the hydrogen atoms and clusters are static, the diffusion coefficient $(D(T))$ and migration energy of hydrogen $(E_M^H)$ mentioned in Juan et al. [32,38] formulation can be neglected. Further, the expression for relaxation time, $\tau$ can be simplified to,

$$\tau = \frac{L_P C_0 (kT)^{\frac{3}{2}}}{3GSb}\exp\left(\frac{2E_K + E_B}{kT}\right) \qquad (17)$$

where $C_0$, G, S are variables and are described in detail in [32]. From the above expression of relaxation time, the effective activation energy associated with SK peak, $E_{SK}$ can be deduced as follows,

$$E_{SK} = 2E_K + E_B - \frac{3}{2}kT \qquad (18)$$

where $2E_K$ represents the activation energy of double kink and $E_B$ is the binding energy for dislocation and hydrogen. The expression of activation energy (Eq. 18) appropriately highlights the contributions of hydrogen on dislocation movement. Interestingly, Eq. (18) is similar to that of Eq. (14) used in the present work for activation energy estimation by considering dislocation-hydrogen interactions.

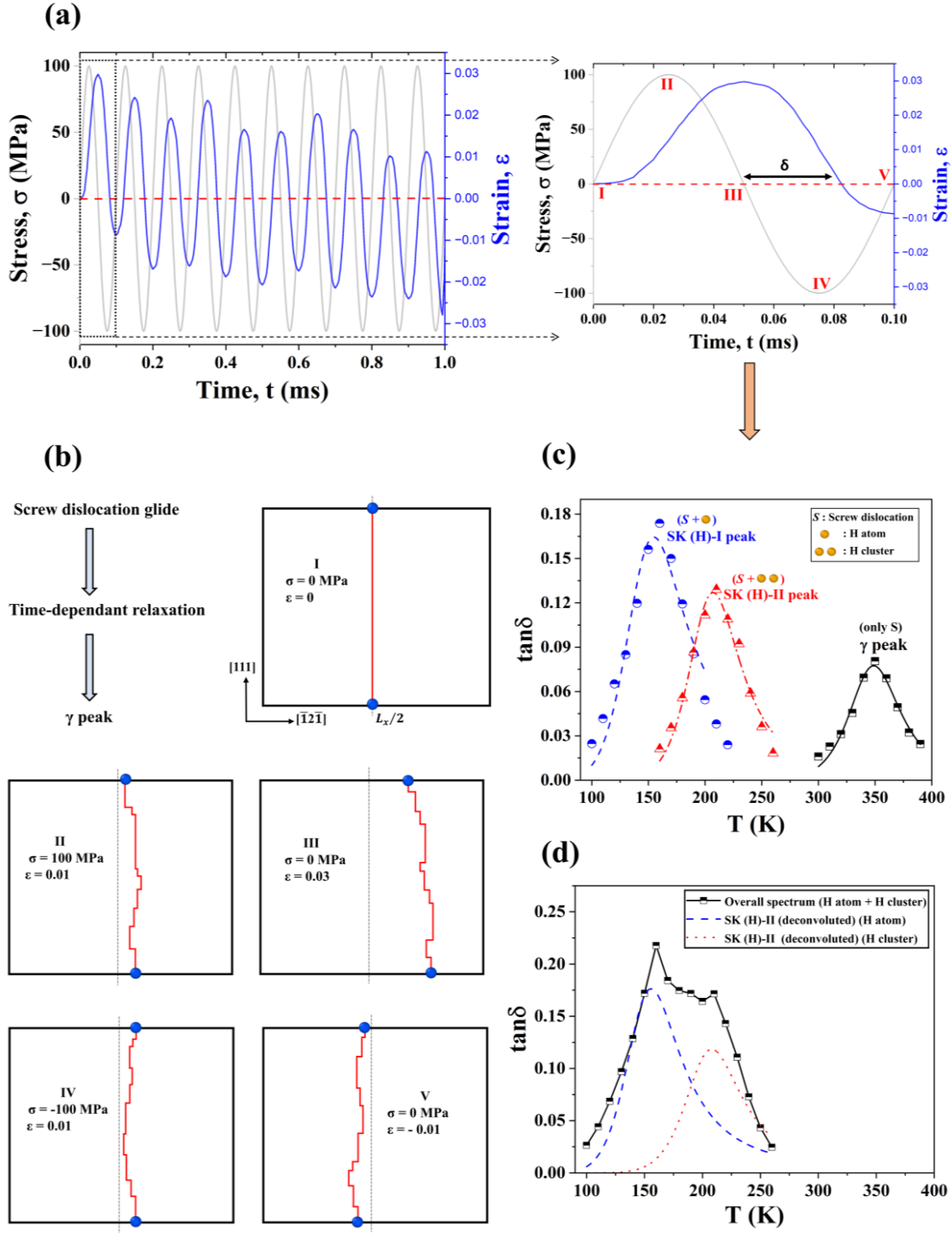

FIG 4. (a) Cyclic stress-strain response captured through KMC simulations of a dislocation movement. (b) Atomistic mechanisms responsible for the $\gamma$ peak. (c) Corresponding internal friction loss factor, $\tan\delta$ versus temperature ($T$) depicting $\gamma$, SK(H)-I and SK(H)-II peaks arising from intrinsic dislocation relaxations, and dislocation interactions with hydrogen atoms and clusters, respectively. (d) Deconvolution of complex internal friction spectrum into SK(H)-I and SK(H)-II peaks.

**D. Influence of Hydrogen Concentration**

Furthermore, the contributions of hydrogen concentration on SK(H) peaks have been evaluated and summarized in Fig. 5. In this context, three varying hydrogen concentrations were considered. It was noticed that internal friction losses increased with hydrogen content for both hydrogen atoms and clusters, see Fig. 5a. The peak height, $\tan \delta_{max}$ showed a linear increase with at% H, as evident from Fig. 5b. Fig. 5 revealed that hydrogen content plays a dominant role in SK(H) peaks. This information can be utilized for hydrogen quantification based on the correlation between $\tan \delta_{max}$ and hydrogen content (at% H). It should be noted that the accurate identification and quantification of hydrogen atoms and clusters through experimental techniques is challenging. Moreover, the relative contributions of hydrogen atoms and clusters to internal friction losses could not be isolated with the experimental observations. However, this limitation was thoroughly addressed by current MD-KMC simulations.

Till now, a few limited studies [33,63] have shown a quantitative trend between internal friction loss factor, $\tan \delta$, and hydrogen content. In particular, the study by Sturges et al. [63], showed a qualitative match with our MD-KMC simulations, as shown in Fig. 5b. Nevertheless, the correlation between $\tan \delta$ and hydrogen content appears to be sensitive to hydrogen type (atom or cluster) as well. For instance, internal friction peaks through dislocation interactions with hydrogen atoms and clusters, SK(H)-I and SK(H)-II, have been decoupled with MD-KMC simulations. Otherwise, this clear distinction is not possible with the experiments due to the complex nature of the microstructures. Moreover, the present simulations demonstrate a linear scaling between hydrogen content (at% H) and $\tan \delta$ for both SK(H)-I and SK(H)-II peaks, see Fig. 5b. This proves the accuracy of the present internal friction simulations for hydrogen quantification. The present simulations were instrumental in decoupling the contributions from hydrogen atoms and clusters to the overall internal friction response. The slopes between $\tan \delta$ and at% H are slightly different for SK(H)-I and SK(H)-II peaks.

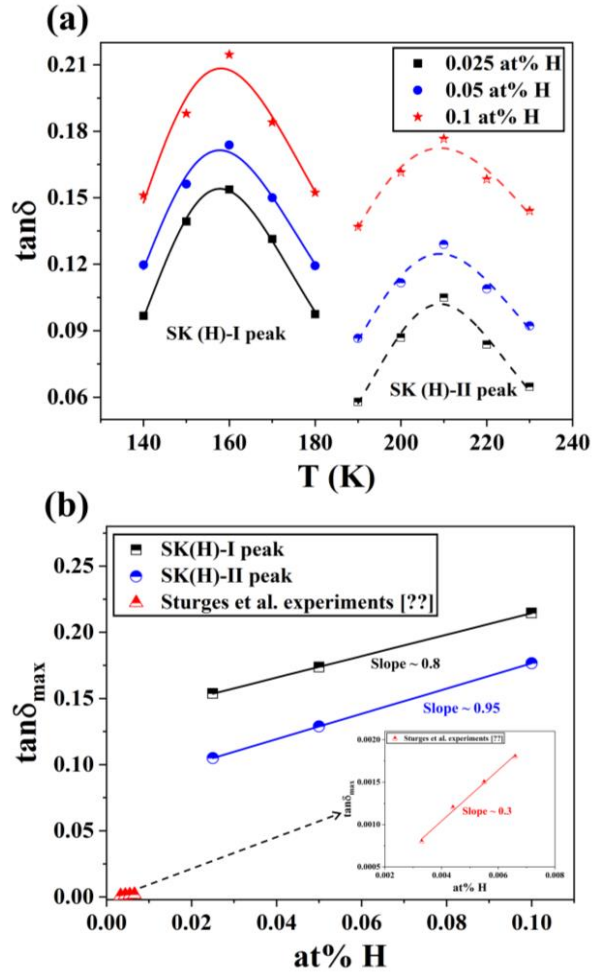

FIG. 5. (a) Simulating SK(H)-I and SK(H)-II peaks for different hydrogen concentrations. (b) SK peak height variation with hydrogen concentration (at% H).

The difference in slopes is from the variation in the fraction of atoms and clusters with a change in hydrogen concentration. As hydrogen concentration (at% H) is varied from 0.025 at% H to 0.1 at% H, hydrogen cluster content increases from 5 at% to 20 at%, while hydrogen atom fraction reduces from 95 at% to 80 at%. As a result, SK(H)-II will dominate with the increase in hydrogen concentration, and this is supported by higher slopes for SK(H)-II in Fig. 5b. This observation validates the hydrogen clustering tendency at higher hydrogen concentrations. Therefore, the variation in the peak height of SK(H)-I and SK(H)-II peaks can be utilized as an efficient method for hydrogen detection, as well as its nature. The present findings can readily be extended to more realistic microstructures and develop a direct correlation between internal friction losses and hydrogen content. Notably, site-specific local internal friction measurements, coupled with hydrogen charging tests and niche microstructural observations, can be strategically designed to enable efficient mapping of hydrogen and its correlation with mechanical performance. Such focused experiments and atomistic simulations are crucial to uncover the underlying causes of hydrogen embrittlement and potentially mitigate them by altering the microstructure and restricting hydrogen diffusion.

**E. Atomistic Mechanisms of Dislocation Relaxations**

The primary motivation of the present study is to reveal atomistic mechanisms associated with internal friction peaks arising from anelastic relaxation of dislocations. In addition, the role of hydrogen-dislocation interactions on internal friction behavior has been investigated. This study effectively decoupled multiple internal friction peaks and correlated them to their respective relaxation mechanisms. This was achieved by a combination of atomistic MD-KMC simulations on a single screw dislocation surrounded by a hydrogen atoms and clusters. This study brought out new perspectives towards dislocation relaxation through internal friction simulations. It has been demonstrated that dislocation relaxation phenomena are heavily influenced by kink nucleation and migration behaviors, which are, in turn, quite sensitive to variations in temperature and applied stress.

More importantly, this study highlighted the role of hydrogen-dislocation interactions on the internal friction response. MD simulations provided the nucleation and migration barriers for dislocation glide in a hydrogen-rich environment. These barriers, as summarized in Fig. 2b-c, were computed by placing selected hydrogen atoms/clusters schematically at different locations around a screw dislocation. While these barrier values were very limited in number, they were sufficient to initiate KMC simulations. To present a broad perspective, additional nucleation and migration barrier values were extracted from multiple KMC simulations conducted over a wide spectrum of temperature and stress levels. It is quite difficult to gather such a large range of interesting data through experimental observations alone and propose underlying mechanisms. Hence, the present simulations proved valuable in developing a link between the observed internal friction spectrum and associated relaxation mechanisms. In particular, stress values spanning the full stress cycle (-100 MPa to 100 MPa) were considered. On the other hand, the temperature range was chosen based on the appearance of particular internal friction peaks. For instance, the $\gamma$ peak was observed in the temperature range of 300-400 K. Whereas, SK(H)-I and SK(H)-II peaks appeared in the 100-200 K and 160-260 K ranges, respectively. These data were obtained accordingly and are presented as ranges in Fig. 6a-b.

It can be observed that the nucleation barrier ($\Delta H_{kn}$) is quite higher than the migration barrier ($\Delta H_{km}$) for pure screw dislocation relaxation, corresponding to the $\gamma$ peak. The observed values of $\Delta H_{kn}$ are in the range of 0.35-0.5 eV, while $\Delta H_{km}$ lies in the range of 3-9×10$^{-6}$ eV.

This large disparity suggests that kink nucleation event is the rate-limiting step in screw dislocation motion, and hence, it governs the overall diffusion and kinetics of a pure screw dislocation in the absence of hydrogen. The high nucleation barrier implies that dislocation glide is difficult and time-dependent. Consequently, pure screw dislocation mobility is extremely low within the temperature range of 100-300 K, as also evident from Fig. 3c. Significant dislocation movement only occurs above 300 K, which explains why pure screw dislocation relaxation appears at relatively high temperatures (300-400 K). In contrast, the presence of hydrogen caused a substantial reduction in the nucleation barrier ($\Delta H_{kn}$), while a significant increase in the values of $\Delta H_{km}$ was noticed. For instance, $\Delta H_{kn}$ corresponding to the SK(H)-I peak is reduced to the range of 0.15-0.38 eV, while $\Delta H_{kn}$ ranged between $0.21 - 0.44$ eV for SK(H)-II peak with significant decrease, as summarized in Fig. 6a. This reduction is due to enhanced hydrogen-dislocation interactions, which lead to the dislocation attraction towards hydrogen Moreover, these prevalent hydrogen-dislocation interactions have been quantified in terms of the interaction energy term using Eq. 13 and 14. The resulting interaction energy effectively lowered the dislocation line energy. This, in turn, caused a large decrease in Peierls barriers for kink nucleation events. Interestingly, the decrease in $\Delta H_{kn}$ was more pronounced for the SK(H)-I peak than for SK(H)-II. This difference originated from a high fraction of hydrogen atoms (around 80-95 %) surrounding the dislocation core in the case of SK(H)-I.

The relatively lower values of $\Delta H_{kn}$, in the hydrogen atmosphere facilitated easier dislocation movement. This resulted in higher dislocation velocity for the Fe-H system compared to the pure Fe system. Due to this enhanced dislocation mobility, hydrogen-assisted dislocation relaxation peaks shifted to relatively lower temperatures. In essence, hydrogen atoms or clusters make dislocations mobile at relatively low temperatures by reducing the activation energy required for dislocation glide, enabling anelastic relaxation responsible for internal friction peaks. Particularly, such hydrogen-assisted peaks, namely, SK(H)-I and SK(H)-II peaks, were found in the temperature ranges of 100-200 K and 160-260 K, respectively. This behavior aligns with the Hydrogen-Enhanced Localized Plasticity (HELP) mechanism [42,43,46], which suggests that hydrogen atoms accumulate near dislocations, reducing the resistance to their movement and thereby enhancing dislocation mobility.

On the other hand, hydrogen has an opposing effect on the migration barriers. The migration barriers ($\Delta H_{km}$) for the SK(H)-I and SK(H)-II peaks are very high in comparison to the $\gamma$ peak. Specially, $\Delta H_{km}$ enhanced from $5 \times 10^{-6}$ to 0.05 eV for the SK(H)-II peak. Whereas, $\Delta H_{km}$ varied from $5\times10^{-6}$ to 0.15 eV for the SK(H)-I peak. This considerable enhancement in migration barriers is attributed to hydrogen-induced pinning or solute drag, which hampers dislocation migration. Since the dislocation core is a region of very high core energy and extreme stress fields, it acts as a strong trap for hydrogen atoms or clusters. The accumulation of hydrogen around the dislocation kinks restricted the migration process, thus increasing the migration barrier. These higher migration barriers and dislocation-pinning imposes additional restrictions on back and forth movement of the screw dislocation in response to imposed cyclic stress. As a result, a delayed response from the dislocation is expected, and this can accumulate over repeated stress cycles. Such delay is a signature of strong attenuations and contributes to higher internal friction losses, $\tan\delta$ which are characteristics of the SK-(H)-I and SK(H)-II peaks.

To identify the origin of enhanced peak heights of SK(H)-I and SK(H)-II, dislocation configurations were analyzed, focusing on the number and size of kinks. The number of kinks for all three peaks are included in Fig. 6c. It was observed that the number of kinks or dislocation segments are in the range of 4-20 for the SK(H) peaks, while this number is quite low for the $\gamma$ peak. The lower nucleation barriers and enhanced dislocation activity corresponding to the SK(H) peaks promoted a greater kink density.

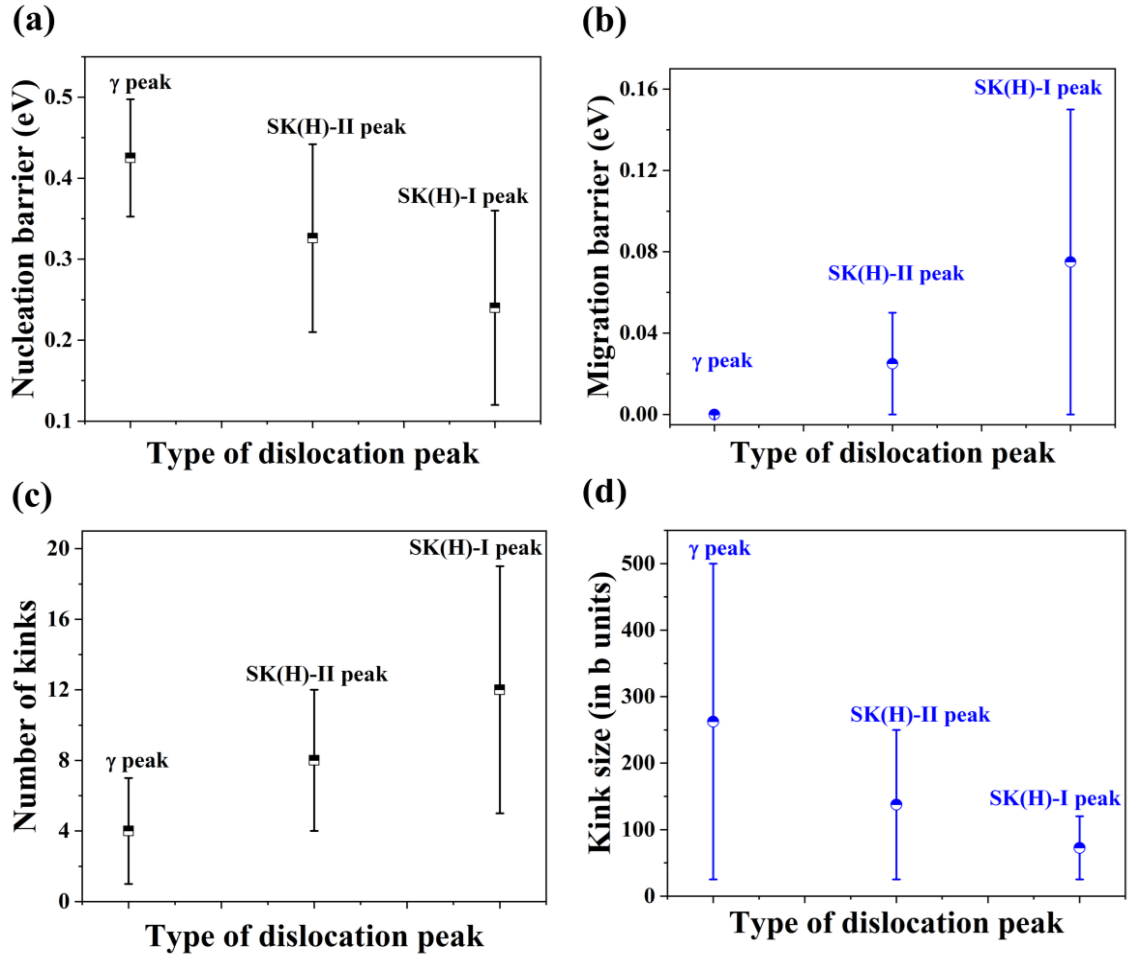

FIG. 6. (a) Nucleation barrier, (b) Migration barrier, (c) Number of kinks, and (d) Kink size for different dislocation peaks, namely, $\gamma$, SK(H)-I and SK(H)-II.

It should be noted that each kink evolves dynamically during the simulation run and contributes towards the overall characteristics of dislocation relaxation. Since SK(H) peaks contained a high number of kinks, the total energy dissipation, tan $\delta$ is too high due to the cumulative effect of energy dissipation from numerous kink events. This effect is enhanced multifold for higher hydrogen concentrations, as demonstrated in Fig. 5. The introduction of hydrogen facilitates the formation of a greater number of kinks, thereby increasing the kink density. This increase is accompanied by a reduction in individual kink size, as is shown clearly in Fig. 6d. The combined statistics of kink size and density significantly influence the internal friction peak height. The outputs of KMC simulations, demonstrating dislocation evolution, were analyzed further to illustrate the underlying mechanism responsible for the SK(H) peaks. In this context, Fig. 7 presents selected KMC snapshots that reveal hydrogen-dislocation interactions. It clearly shows that hydrogen promotes kink nucleation by lowering the nucleation energy barrier, thereby promoting the formation of a higher number of kinks or dislocation segments. These small segments subsequently undergo rearrangement according to their characteristic timescales and the imposed frequency during simulations. Interestingly, the role of hydrogen on the migration barrier is opposite to that of nucleation. The migration process is limited by the higher hydrogen concentration near the kink edges. This led to the dislocation pinning or solute drag effect, attributed to the formation of a hydrogen Cottrell atmosphere. So, this dual behavior of hydrogen, promoting kink nucleation while hindering migration, is key to the emergence of the SK(H) peaks at relatively lower temperature and with higher peak intensity than the $\gamma$ peak. Notably, this mechanism can be extended to other solute atoms (such as C, N, or O) by incorporating the appropriate interaction parameters.

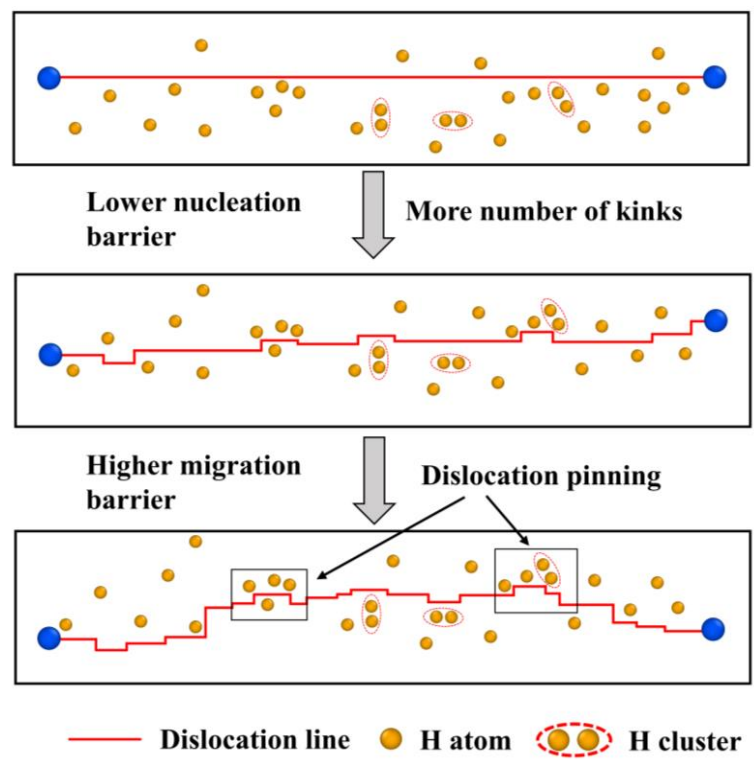

FIG. 7. Overall mechanism responsible for the SK (H) peaks.

Most of the existing theories concerning SK peaks are based on solute drag phenomena [33,64]. This drag effect can be manifested by combining solute diffusion (hydrogen, in this case) and dislocation glide, simultaneously. The solute has two major roles: first, the modification of the intrinsic properties of dislocations by the presence of the solute or Cottrell atmosphere; and second, the physical dragging of solute atoms along the dislocation movement direction. The present simulations incorporated the hydrogen-dislocation interactions and evaluated their impact on dislocation glide events. However, the explicit inclusion of the physical solute drag effect was not feasible within the current KMC simulation framework. The limitations originated from the large separations in the rates of these two key events, i.e., hydrogen diffusion and dislocation glide, also summarized in Fig. 3d. This means that hydrogen diffuses so rapidly compared to dislocation movement that it can be considered static with respect to dislocation evolution. Therefore, solute drag is unlikely to significantly influence the overall relaxation. Moreover, hydrogen atoms are placed randomly over the entire system, which maintains a uniform Cottrell atmosphere around the moving dislocation core. Thus, only dislocation movement behavior was explicitly modeled. With this elegant approach, a lot of simulation time was saved, and still captured all necessary features of the SK peaks.

In summary, this study comprehensively addressed the key characteristics of screw dislocation glide, encompassing diffusion barriers, hydrogen-dislocation interactions, dislocation mobility, and time-dependent relaxations responsible for various internal friction peaks. The findings revealed novel atomistic mechanisms of anelastic relaxation of a screw dislocation under different testing and environmental conditions. The KMC framework employed in the current work is very generalized and simple, making it readily extendable to edge, mixed dislocations, or even more complex dislocation configurations. While such extensions lie beyond the purview of the present theme, they offer promising directions for future research. Any expansion would, however, require suitable modifications based on the specific system and variables involved. Strikingly, a linear correlation between $\tan\delta_{max}$ and hydrogen concentration was uncovered. This suggests that internal friction measurements could serve as a promising tool to estimate the hydrogen content precisely. In addition, this insight may help in addressing a few 'unknowns' associated with the 'hydrogen embrittlement' field. Overall, the present study provides an atomistic-level understanding of the mechanisms driving dislocation relaxation peaks and proposes a potential technique for hydrogen detection.

## IV. CONCLUSIONS

In summary, a unified approach consisting of MD-KMC simulations was proposed to bring the atomistic origin of dislocation relaxation peaks, in particular SK peaks. Notably, the dual role of hydrogen on dislocation diffusion barriers was quantified efficiently. The MD simulations revealed that the kink nucleation barrier ($\Delta H_{kn}$) diminishes in the presence of hydrogen atoms in the vicinity of the dislocation core. In contrast, the migration barriers ($\Delta H_{km}$) were increased by enhanced hydrogen-dislocation interactions. Consequently, Cottrell atmospheres formed by the hydrogen around the dislocation improved dislocation mobility substantially, leading to the emergence of SK peaks at relatively lower temperatures. The relaxation peak strengths of the SK peaks increased drastically due to a higher density of the kinks and additional anelastic sources compared to the $\gamma$ peaks. More importantly, the underlying mechanisms behind SK(H) peaks were elucidated with a minimalistic model based on the diffusion barriers. The present MD-KMC simulations were successful in isolating the contributions of hydrogen atoms and clusters on the overall SK spectrum, by capturing two separate peaks, namely SK(H)-I and SK(H)-II. Such a distinction is otherwise not feasible, experimentally. Especially, a linear scaling between internal friction peak height ($\tan \delta_{max}$) and hydrogen concentration has been observed. This correlation can be exploited further to develop a potential tool for hydrogen detection and quantification at the microscopic level. In the future, the output from the current MD-KMC setup can be integrated with continuum frameworks to simulate more realistic dislocation dynamics and tackle more complex real-world problems.

# Supplementary Information for:
# Influence of Hydrogen on Dislocation Relaxation in BCC Iron: An Internal Friction Perspective

Sanjay Manda[1,$], Madhur Gupta[1,$], Saurabh Kumar[1], Junaid Akhter[1], P. J. Guruprasad[2], Indradev Samajdar[1,*], and Ajay S. Panwar[1,*]

[1] Department of Metallurgical Engineering and Materials Science, Indian Institute of Technology Bombay, Mumbai, 400076, India.

[2] Department of Aerospace Engineering, Indian Institute of Technology Bombay, Mumbai, 400076, India.

[$] Joint first co-authors.

## S1: Implementation of NEB Method

The MD simulations were performed using the package with the interatomic EAM potential developed by Proville et al. [1]. The initial and final configurations were generated using ATOMSK software [2] with a dislocation placed at the center of the box for the initial configuration and at a distance of $\frac{\sqrt{6}}{3}a$ from the box center for the final configuration. These configurations were subsequently relaxed using the conjugate gradient (CG) scheme. The intermediate replicas were linearly interpolated between the relaxed initial and final configurations. The Peierls barrier was computed using the NEB method, which enables the search for the minimum energy path between initial and final configurations. Once all the replicas are relaxed towards the minimum energy path, the replica with the highest energy is relaxed towards the saddle point [3,4]. All the interatomic replicas are subjected to two force vectors: (i) parallel force, which is responsible for maintaining equal spacing between replicas, and (ii) perpendicular force, which prevents the paths from forming acute kinks along the transition path. For this work, the parallel and perpendicular spring constants were set to 10 eV/Å. The NEB was performed with zero force tolerance and an energy tolerance of 0.01.

## S2: Calculation of Migration Barriers

For dislocation migration, the interaction energy of the dislocation and hydrogen ($\Delta \tilde{E}_p$) can be given by:

$$\Delta \tilde{E}_p = \sqrt{\sum_{ij} c \, \Delta E_{sol}(a)^2} \qquad (S1)$$

where $E_{sol}(a)$ represents the change in interaction energy, and c is the solute concentration.

The characteristic stress ($\tau_c$) can be expressed as,

$$\tau_c = \frac{\Delta \tilde{E}_p}{ab^2} \qquad (S2)$$

where a and b are the lattice parameters and Burgers vector, respectively. The enthalpy barrier for kink migration barrier can be given by:

$$\Delta H_{km} = \Delta \tilde{E}_p \left[ 3.26 \left( \frac{\tau}{\tau_c} + \frac{2.7}{\sqrt{L/b}} \right)^{-1} + 0.035\sqrt{L/b} - 1.07\sqrt{w/b} \right] \qquad (S3)$$

Here, $\tau$ is the applied stress, L is the length of the dislocation, and w is the kink width. For further details regarding this model, please refer to [5].

## S3: Diffusion Barrier Estimation for Fe-H system

Fig. S6 presents the minimum energy path for dislocation motion from one Peierls valley to another in the Fe-H system, as predicted using MD simulations. The well-established Proville EAM potential [1] was employed to capture the minimum energy path for a pure Fe system, as it accounts only for Fe-Fe interactions. For Fe-H interactions, we used the EAM potentials proposed by Kumar et al. [6] and Wen et al. [7]. However, the figure reveals that neither potential accurately predicts the minimum energy path for dislocation-hydrogen interactions. The Kumar EAM potential [6] results in an overestimated peak height, whereas the Wen EAM potential [7] produces a double-hump profile. Since these individual potentials fail to provide accurate results, we employed a hybrid approach: the Proville EAM potential was used for Fe-Fe interactions, while the Kumar/Wen potentials were applied for Fe-H interactions. Despite this refinement, the minimum energy path still overestimates the dislocation-hydrogen interactions, as shown in Fig. S6. Consequently, to accurately determine the activation energy barrier for dislocation-hydrogen interactions, we adopted an alternative method that explicitly incorporates the contribution of solute atoms in terms of interaction energy. Further details of this formulation are provided in Eq. 13 of the main manuscript.

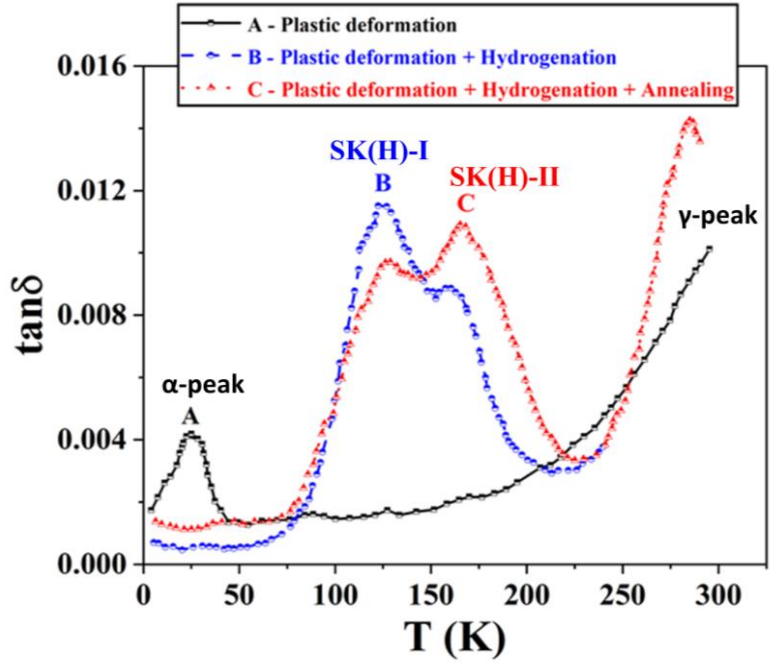

FIG. S1. Internal friction measurements on pure iron demonstrated the presence of various dislocation relaxation peaks, namely, $\alpha$, $\gamma$ and SK(H) peaks. These are shown through multiple experiments. Here, the plot marked as A belongs to the plastic deformation sample at room temperature, while plot B was extracted after hydrogenation of the same sample and plot C was obtained after an additional plastic deformation at 85 K and subsequent annealing at 265 K. This figure is adopted from San Juan et al. work [8].

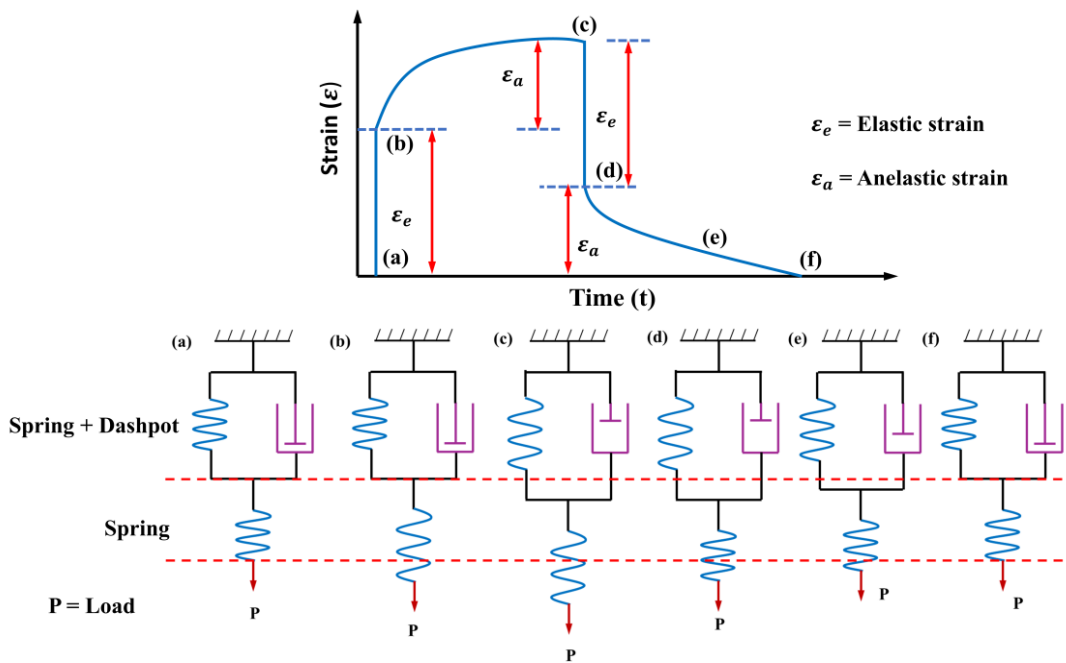

FIG. S2. Anelastic behavior of a solid is demonstrated through the Voigt-spring model. In this model, a spring and a dashpot are arranged in a series and attached to another spring. The corresponding strain response is shown along with different stages of the Voigt-spring model.

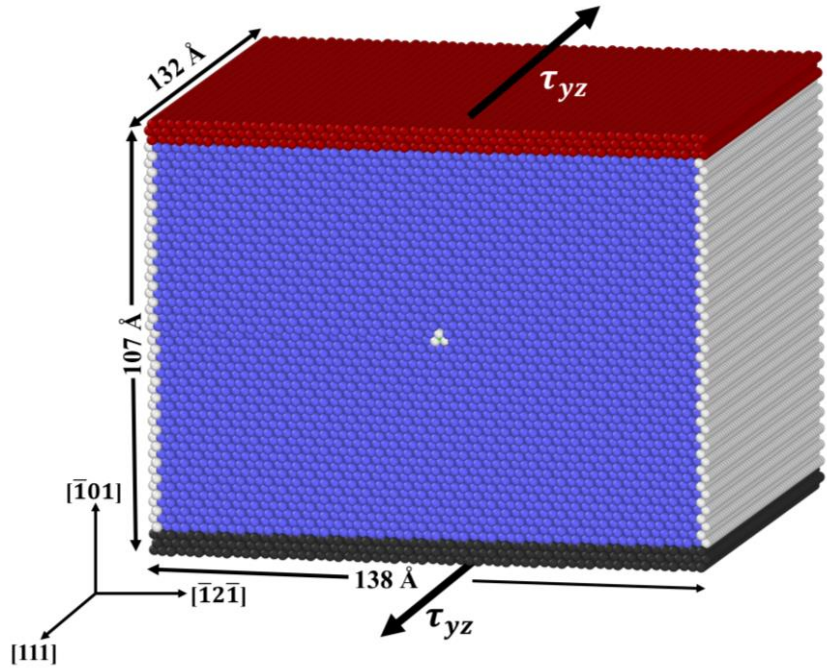

FIG. S3. Simulation setup containing a screw dislocation was used in the present simulation. Simulation box dimensions, axes convention, and shear stress direction are also included.

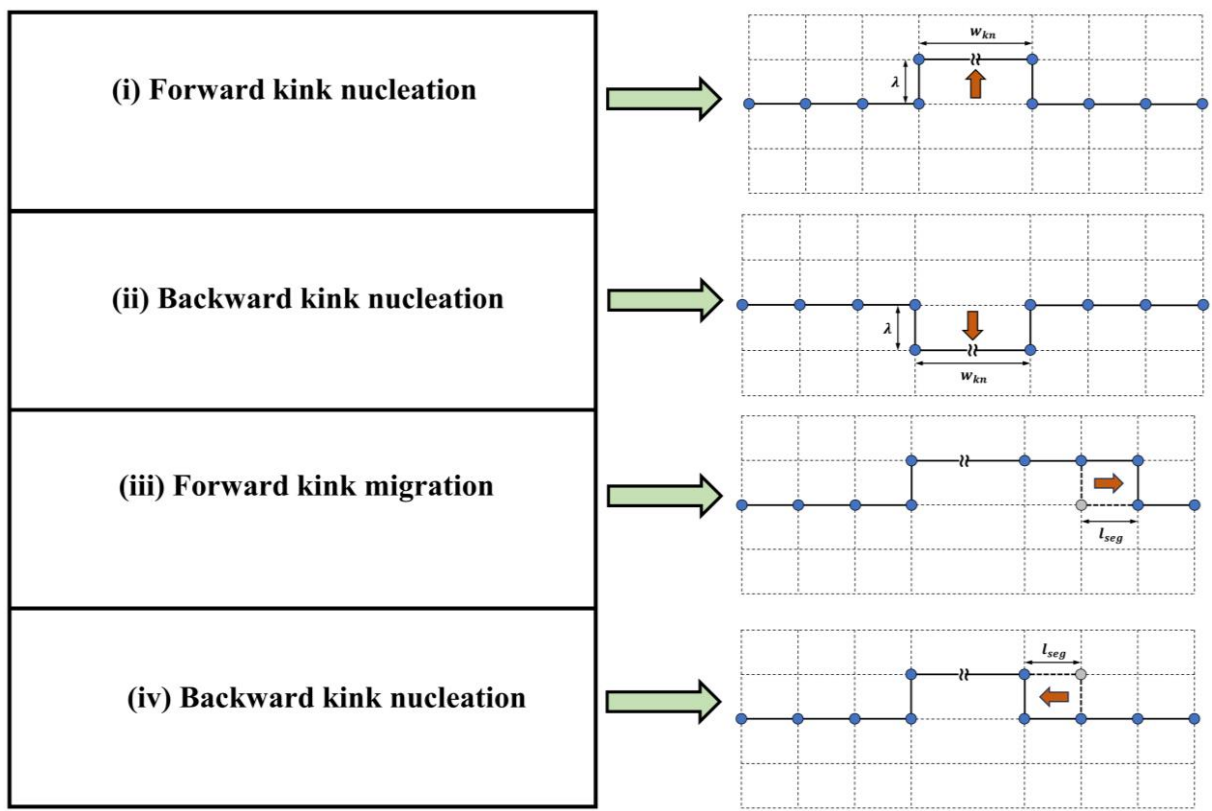

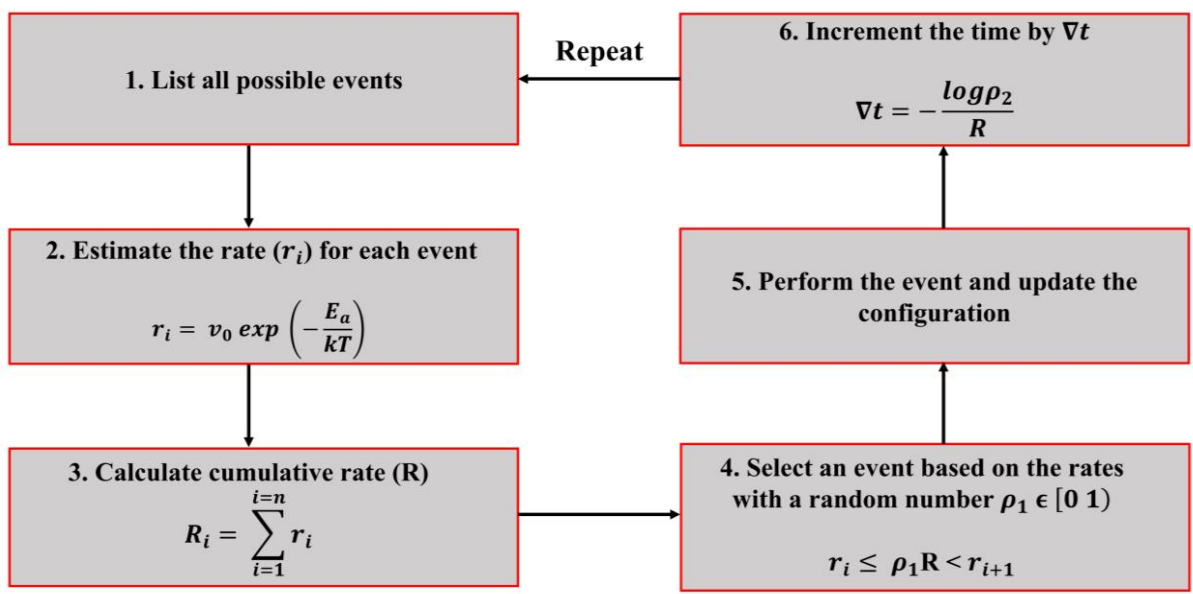

FIG. S4. (a) Four major events belonging to the dislocation glide process are the following: (i) Forward kink nucleation, (ii) Backward kink nucleation, (iii) Forward kink migration, and (iv) Backward kink migration. (b) The details of the n-fold way algorithm.

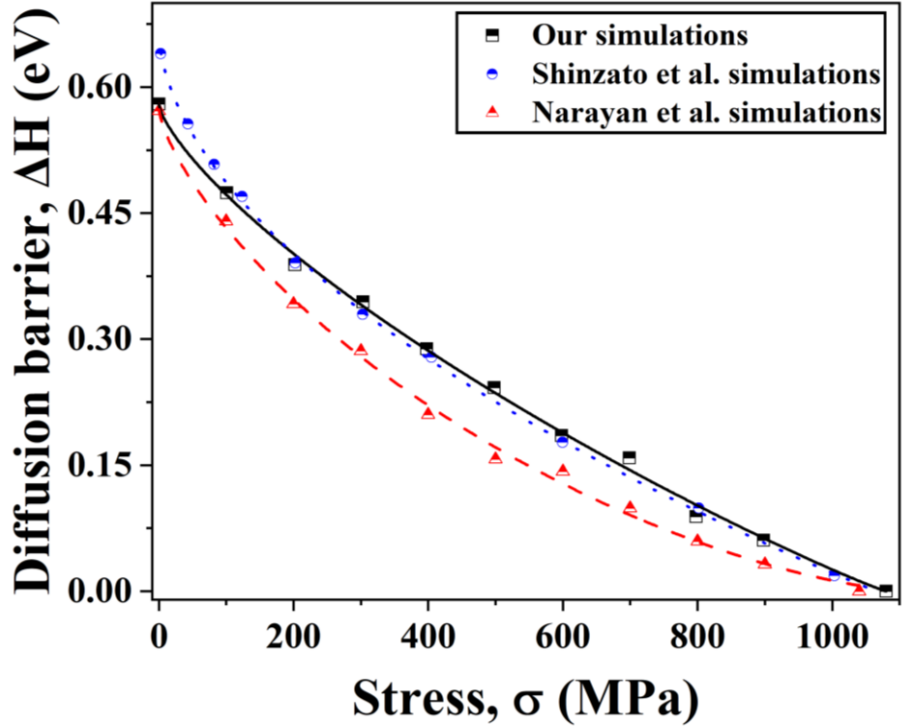

FIG. S5. Variation of maximum height of energy landscape (diffusion barrier, $\Delta H$) as a function of applied shear stress. These values are validated against literature values reported by Shinzato et al. [9] and Narayan et al. [10].

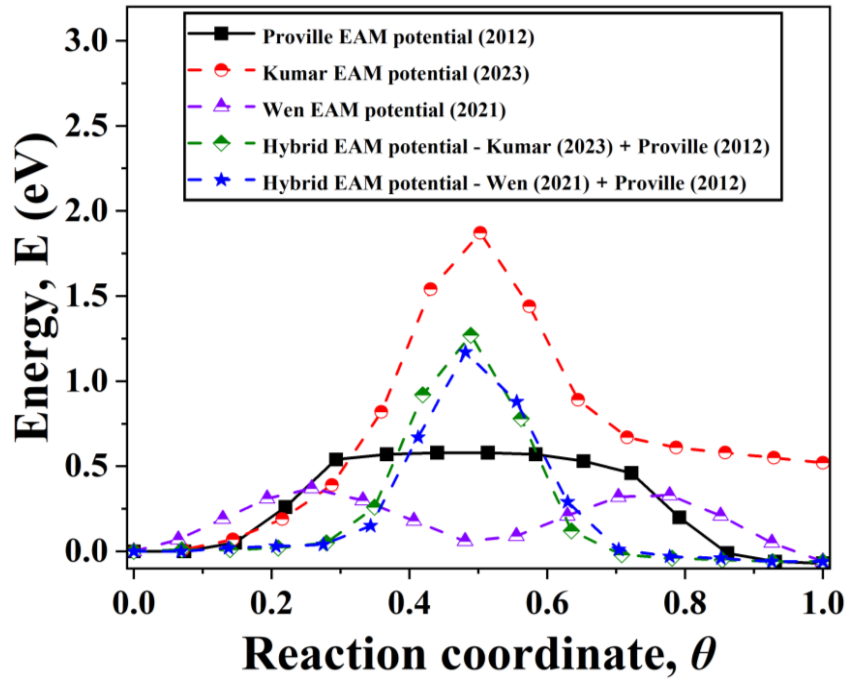

FIG. S6. Calculation and validation of diffusion barrier ($\Delta E$) for Fe-Fe and Fe-H systems by using multiple EAM potentials.

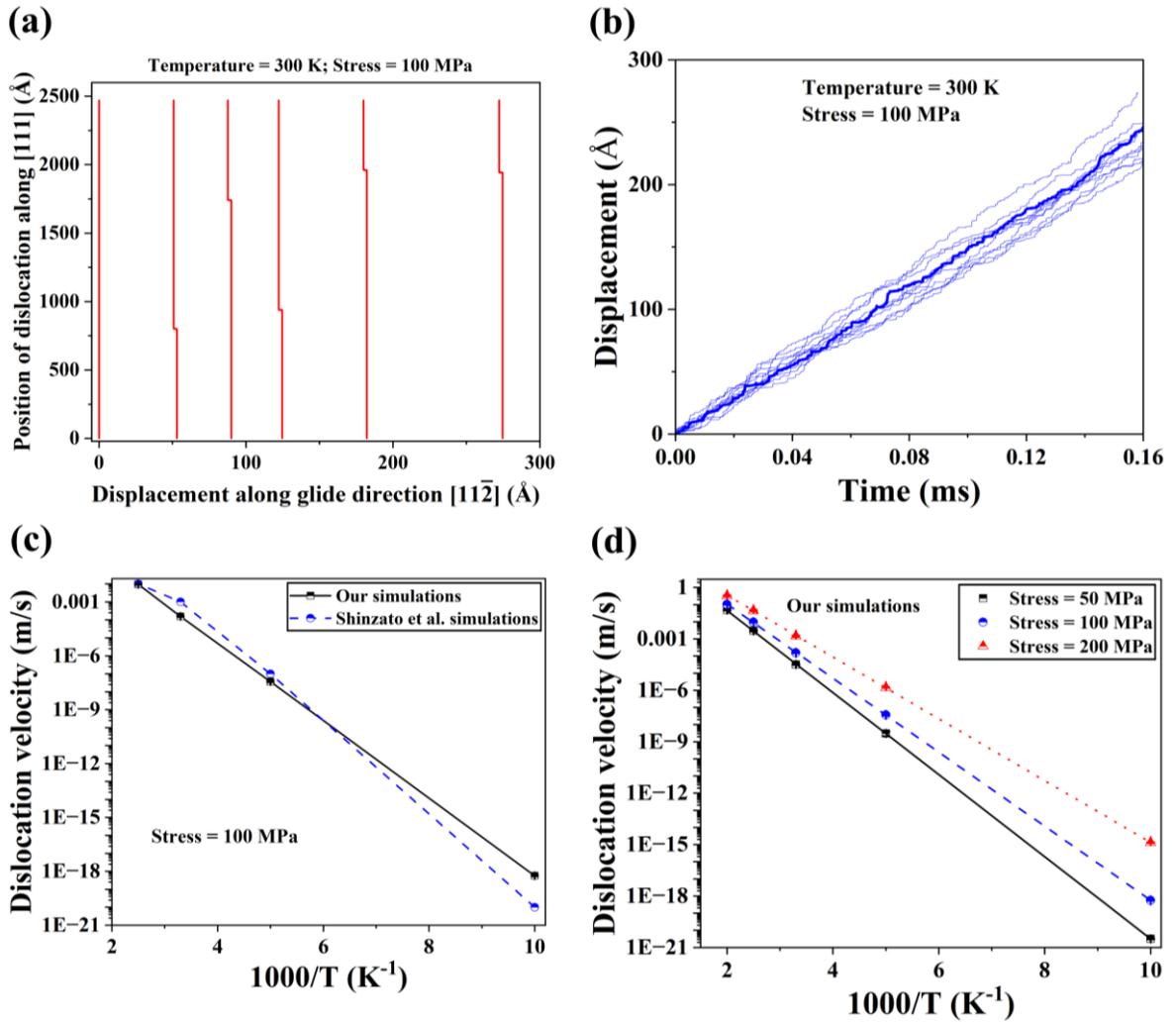

FIG. S7. (a) Different stages of a screw dislocation movement through kink nucleation and migration events while maintaining the connectivity between various segments. (b) Dislocation displacement over a time scale. (c) Dislocation velocity as a function of temperature validated against literature data. (d) Dislocation velocity at different values of imposed constant stress.

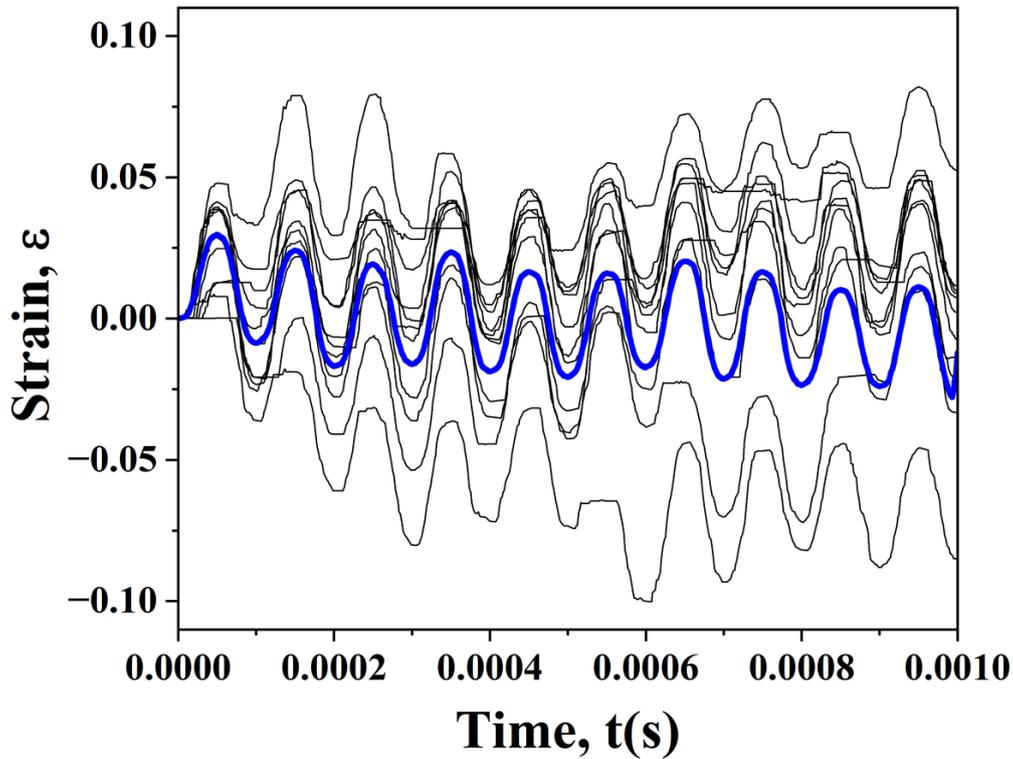

FIG. S8. KMC simulated multiple strain trajectories are plotted against simulation time along with the average strain.